\documentclass[reprint,superscriptaddress,unsortedaddress,amsmath,amssymb,aps,prl]{revtex4-1}

\usepackage{graphicx}
\usepackage{dcolumn}
\usepackage{bm}
\usepackage{xcolor, ulem}
\usepackage{comment}
\usepackage{hyperref}

\begin{document}

\title{Dilute suspensions of Janus rods: the role of bond and shape anisotropy}
\author{Carlo Andrea De Filippo}
\affiliation{Science Department, University of Roma Tre, Via della Vasca Navale 84, 00146, Rome, Italy}
\author{Sara Del Galdo}
\affiliation{Science Department, University of Roma Tre, Via della Vasca Navale 84, 00146, Rome,  Italy}
\author{Emanuela Bianchi}
\affiliation{Institut f{\"u}r Theoretische Physik, TU Wien, Wiedner Hauptstra{\ss}e 8-10, A-1040 Wien, Austria and CNR-ISC, Uos Sapienza, Piazzale A. Moro 2, 00185 Roma, Italy}
\author{Cristiano De Michele}
\affiliation{Physics Department, University of Roma ``La Sapienza'',Piazzale Aldo Moro 2, 00186,  Rome, Italy}
\author{Barbara Capone}
\affiliation{Science Department, University of Roma Tre, Via della Vasca Navale 84, 00146, Rome,  Italy}
\date{\today}

\begin{abstract}
Nanometer-sized clusters are often targeted due to their potential applications as nanoreactors or storage/delivery devices. One route to assemble and stabilize finite structures consists in imparting directional bonding patterns between the nanoparticles. When only a portion of the particle surface is able to form an inter-particle bond, finite-size aggregates such as micelles and vesicles may form. Building on this approach, we combine particle shape anisotropy with the directionality of the bonding patterns and investigate the combined effect of particle elongation and surface patchiness on the low density assembly scenario. To this aim, we study the assembly of tip-functionalised Janus hard spherocylinder by means of Monte Carlo simulations. By exploring the effects of changing the interaction strength and range at different packing fractions, we highlight the role played by shape and bond anisotropy on the emerging aggregates (micelles, vesicles, elongated micelles and lamellae). We observe that shape anisotropy plays a crucial role in suppressing phases that are typical to spherical Janus nanoparticles and that a careful tuning of the interaction parameters allows to promote the formation of spherical micelles. These finite-size spherical clusters composed of elongated particles might offer more interstitials and larger surface areas than those offered by micelles of spherical or almost-spherical units, thus enhancing their storage and catalytic properties.
\end{abstract}

\maketitle

Colloidal design has gained a prominent role in the realisation of novel functionalised materials with a plethora of potential applications, such as in photonics, food industry, drug delivery, and energetics~\cite{Nanolett_2011,osada2012two,CSR_2014,Nanoscale_2014,ChemComm_2015,CR_2016,mendoza2016synthesis,Nanotoday_2018}. The continuous advancement of experimental techniques has led to the synthesis of nanoparticles with well-controlled shape and fine-tuned functionalisation~\cite{pawar,Bianchi2017Limiting,ravaine2017synthesis,zhang2004self,manoharan,sacanna2013shaping}. The resulting anisotropic nature of the inter-particle interactions is responsible for the spontaneous formation of complex structures, ranging from finite clusters -- mostly in the low temperature, low density regime -- to crystal phases~\cite{han,Wang,Kraft,Tikhomirov,chen,he2020colloidal}. 

Equilibrium finite assemblies at the nanoscale, such as micelles, vesicles, and elongated clusters, are exploited in a wide variety of fields, ranging from biomedicine to nanotechnology~\cite{Ryan,Lin,Bakshi,Khullar2013}. Stable clusters of finite size can in fact act as drug delivery vehicles thanks to their ability to self-assemble and encapsulate substances, while they can also be used as templates for nanoparticle synthesis or even as hosts for chemical and physical reactions at the nanoscale~\cite{Patra,Beverina,Lipshutz,Kim,Ma,Zhao}. 

The spontaneous formation of stable aggregates with a finite size often emerges from the interplay between attractive and repulsive forces at the level of the self-assembling units: while attraction leads a dilute system toward aggregation, repulsion poses challenges to the growth of a space-filling phase. The resulting finite structures have a range of shapes and sizes depending on the nature of the assembling units and the thermodynamic conditions at which they form. 

Drawing inspiration from mechanisms widely observed in polymer systems -- where di- and tri-block copolymers often form micelles, vesicles and elongated finite clusters~\cite{Denkova2009, Sivokhin2021,Mirsaidov,Ooya} -- colloidal particles can be engineered to target finite size aggregates.

Enforcing anisotropic interactions between the self-assembling units is one of the most promising tools to stabilize finite size clusters~\cite{pawar,Bianchi2017Limiting,sacanna2013shaping,zhang2004self,han,Wang,Kraft,Karner,Long,Williamson}. Amongst the class of functionalised colloids, spherical Janus functionalised nanoparticles have been heavily studied in the dilute regime revealing the spontaneous formation of spherical and elongated micelles, vesicles together with wrinkled sheets and different crystal structures at low and intermediate temperatures~\cite{Sciortino2009, Sciortino2010, Sciortino2013, Sciortino2014, Vissers2013, Sciortino2016, Van2017, Sato2020, Li2012, Zou2016, Diaz2019}. 

A step further in complexity is represented by Janus dumbbells~\cite{Whitelam2010, Munaò2014, Dijkstra2015, snowman, Sato2021}: at low densities, non-overlapping patchy dumbbells tend to self-aggregate, forming finite clusters out of the homogeneous fluid when temperature decreases; their phase behavior sensitively depends on the aspect and size ratio of the dumbbells as well as on the strength of attractive interactions, where all three parameters - together with the thermodynamic ones -- may be tuned to promote the development of spherical clusters, i.e. micelles or vesicles, versus elongated micelles. 

It must be noted that, when the anisotropy of the bonding pattern is enhanced even more by anisotropy in the particle shape an even richer assembly scenario unveils. Indeed, shape anisotropy can induce by itself the formation of liquid crystalline mesophases, thus providing a very rich phase behavior\cite{CADF}. Further addition of attractive anisotropic interactions between particles can lead to an even more complex and rich phase behavior due to the interplay between anisotropic attractions and anisoptropic steric repulsion~\cite{Nguyen2014}. Janus dumbbells~\cite{Whitelam2010, Munaò2014, Dijkstra2015, snowman, Sato2021}, rods~\cite{Tripathy2013, Zhang2015, Jurasek2017, Menegon2019, Kobayashi2020, Hossain2022}, and ellipsoids~\cite{Xu2015,Amadei2018} have been often studied in the context of responsive materials; in particular, they have been designed so far to aggregate in chains and non-spherical extended clusters that can rearrange under an external stimulus, provided for instance by electric fields, temperature changes or ultrasounds~\cite{Wolters2017, Kang2018, Liu2022, Yan2013, Zhao2016, Zhang2023, Aayush2013, Chaudhary2012, Murphy2016, Peng2014, Ahmed2014, Oh2019}. Within the broad variety of assembly behaviors observed so in systems of anisotropic Janus colloids, we focus on the emergence of finite clusters. 
In contrast to micelles and vesicles made from spherical colloid or colloidal dumbells, stable finite clusters composed of more elongated objects may open up opportunities for their use as storage/delivery nano-devices as well as nanoreactors \cite{Sorhie2022, Tevet2021, Suehiro2022, Sombat2023}.
These clusters might posses in fact a less densely packed outer layer, thus offering a greater abundance of interstices and a larger surface area compared to micelles formed by spherical or dumbbell-shaped colloids. 

In our work, we consider tip-functionalised Janus-like hard spherocylinders interacting via a Kern-Frenkel potential~\cite{KernFrenkel2003} as prototypes of elongated patchy particles. We employ Monte Carlo simulations to investigate the self-assembly properties of these units: we particularly focus on the low packing fraction regime, specifically targeting the assembly of finite-size objects as a function of particle elongation, packing fraction, interaction strength and interaction range.

A notable feature of our model is that we can tune independently particle patchiness and elongation. By tuning the range and, independently, the strength of the localised interactions, we aim at unveiling how the competition between entropy and enthalpy affects the emerging aggregates (spherical and elongated micelles, vesicles and lamellae), focusing specifically on how to promote or suppress determined structures.
Interaction range and strength are, in fact, parameters used to model the chemical functionalisations on the particles; spanning over a wide range of parameters  allows to explore how  a vast variety of  possible experimental functionalisations would affect the assembly scenario. This is reminiscent to what is done in polymeric systems where tuning interaction parameters to experimental data allows to explore the self assembly properties of macromolecular building blocks \cite{Biopolymers2018,Poleto2022,Chemrev2018}.

\begin{figure}[!h]
    \centering
\includegraphics[width=0.95\columnwidth]{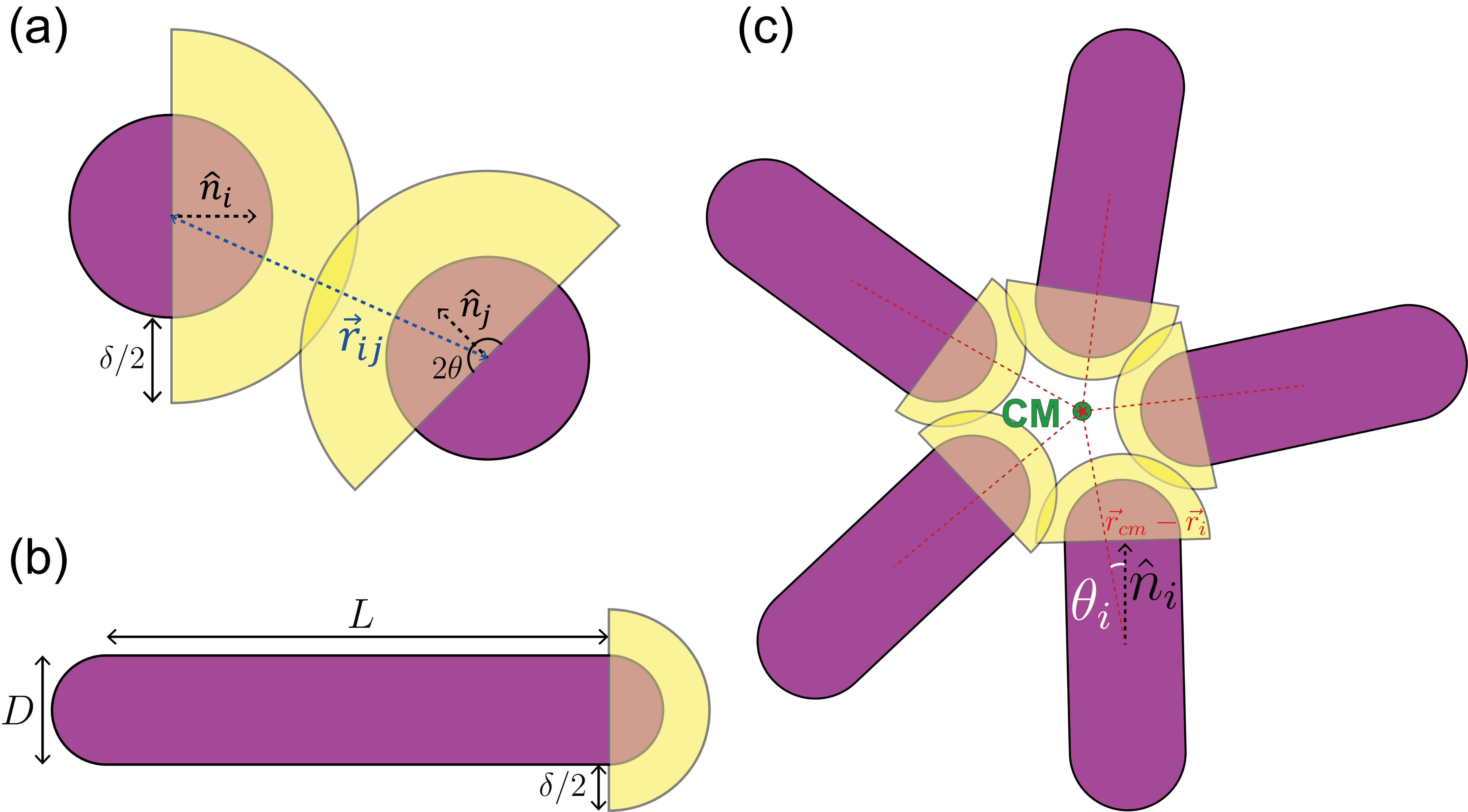}
    \caption{Panel (a): sketch of two interacting Janus spheres with Kern-Frenkel patches, the interaction range is $\delta$ and the patch opening angle $2\theta=\pi$; $\hat{n}_i$ and $\hat{n}_j$ are the orientation unit vectors of the patches on the particles $i$ and $j$, respectively, while $\vec{r}_{ij}$ is the center-to-center vector between the patches. Panel (b): sketch of one Janus-like HSC with elongation $L$, diameter $D$ and a K-F patch with interaction range $\delta$. Panel (c): sketch  of a cluster composed of five rods, where $\Vec{r}_{\mathrm{cm}} - \Vec{r}_i$ is the distance vector between the center of mass of the cluster and the center of the particle $i$, $\theta_i$ is the angle between the axis of the particle (identified by the vector $n_i$) and the vector $\Vec{r}_{\mathrm{cm}} - \Vec{r}_i$.}
    \label{fig:Particle_model}
\end{figure}

It is worth noting that, despite Janus dumbells already introduce some shape anisotropy in addition to the bond anisotropy, the enthalpic and entropic contribution to their assembly  are there intertwined; this is because, for a given a size ratio of the two spheres composing the dumbbells, the sphere separation -- that drives the assembly towards one or another type of aggregates -- affects both the Janus balance and the size asymmetry~\cite{Whitelam2010, Munaò2014, Dijkstra2015, snowman}.

We model Janus rods as hard spherocylinders (HSCs), where one of the tips is decorated with an attractive patch  (see Fig. \ref{fig:Particle_model}). HSCs are defined  by their elongation $L$, their diameter $D$, and the aspect ratio (also referred to as shape anisotropy) $A=L/D$. The tip functionalisation is modelled as an attractive patch through a Kern-Frenkel potential~\cite{KernFrenkel2003} (K-F). The K-F interaction between the particles $i$ and $j$ is given by $U^{o}_{ij}(\hat{n}_i, \hat{n}_j, \hat{r}_{ij}) \cdot U^{d}_{ij}(r_{ij})$, where $\hat{n}_i$ and $\hat{n}_j$ are the orientation unit vectors of the patches on particle $i$ and $j$, respectively, and $\vec{r}_{ij}$ is the center-to-center vector between the patches. $U^{o}_{ij}(\hat{n}_i, \hat{n}_j, \hat{r}_{ij})$ represents the orientational dependence of the potential and reads as
\begin{equation}
    U^{o}_{ij} =
    \begin{cases}
      1 & \text{if } \hat{n}_i \cdot \hat{r}_{ij} > 
      \cos(\theta) \text{ and } \hat{n}_j \cdot \hat{r}_{ij} < - \cos(\theta)\\
      0 & \text{else}
    \end{cases}       
\end{equation}
where $2 \theta$ defines the angular amplitude of the patch (see Fig. \ref{fig:Particle_model}). $U^{d}_{ij}(r_{ij})$ is a square well interaction and reads as
\begin{equation}
    U^{d}_{ij}(r_{ij}) =
    \begin{cases}
      \varepsilon & \text{if } r_{ij} \leq D + \delta \\
      0 & \text{else}
    \end{cases}       
\end{equation}
where the parameter $\varepsilon$ represents the interaction strength and $\delta$ is the interaction range.

We studied systems of $N=2500$ particles, each of volume $v_0$, in a volume $V$ by performing NVT Monte Carlo simulations by adapting the Aggregation Volume Bias algorithm as in Refs.~\cite{Chen_AVB, Chen_AVB_improved, Rovigatti2018}), and discussed in  section 1 of the SI.
In order to identify the assembly scenarios as a function of experimentally accessible parameters, we varied the packing fraction $\phi = N v_0/V \in [10^{-3}, 0.2] $, $A = (0, 0.5, 1.5, 2.5, 3.5, 4.6)$, $\varepsilon= (-2, -3, -4) k_\mathrm{B} T$ (with $k_\mathrm{B}$ the Boltzmann constant), and $\delta/2 = (0.05, 0.25, 0.4) D$. To model the Janus-like K-F interaction, we fixed $\theta = \pi/2$. For each set of $\{\phi,A,\varepsilon,\delta\}$ we performed 2 independent runs.
We collected at least a minimum of twenty equilibrated configurations from each run. Systems were considered to be equilibrated when only negligible drifts of both the internal energy of the system and the observables computed within this work (parameters describing the clusters, \textit{vide infra}) could be detected.
In the AVB-MC we set 0.1 as the probability of the biased move which could be either a bonding (AVB-B) or an unbonding (AVB-U) trial move with equal probability.

We performed a cluster analysis based on an energy criteria: when two particles have a negative pair energy then they belong to the same cluster and a cluster is defined as a set of bonded particles.
Then, to classify the clustered structures we combined a cluster analysis with the characterization of the specific features of the clusters. Hence, once clusters are identified, we build three order parameters that are able to effectively classify clusters based on their microscopic structure~\cite{Dijkstra2015, Van2017, Diaz2019}:

\begin{equation} \label{eq:M}
    \mathcal{M} = \frac{1}{N_c} \sum_{i = 1}^{N_c} cos \theta_i
\end{equation}
\begin{equation} \label{eq:V}
    \mathcal{V} = \frac{1}{N_c} \sum_{i = 1}^{N_c} (1 - \sin \theta_i)
\end{equation}
\begin{equation} \label{eq:B}
    \mathcal{B} = \frac{2}{N_c (N_c - 1)} \sum_{(ij)} (\hat{n}_i \cdot \hat{n}_j)^2
\end{equation}
where $N_c$ is the number of particles composing the cluster (cluster size) and $\sum_{(ij)}$ denotes the non-repetitive sum over the particles in the cluster. The angle $\theta_i$ is obtained from
\begin{equation} \label{eq:cos_theta}
    \cos \theta_i = \hat{n}_i \cdot \frac{\Vec{r}_{\mathrm{cm}} - \Vec{r}_i}{| \Vec{r}_{\mathrm{cm}} - \Vec{r}_i |}
\end{equation}
where $\Vec{r}_{\mathrm{cm}}$ defines the position of the center of mass of the cluster and $\Vec{r}_i$ the position of the center of the $i$-th particle (see panel (c) of Fig. \ref{fig:Particle_model}).
Note that we calculate the aforementioned quantities for all clusters with size larger or equal to four, meaning that we do not include dimers nor trimers in the following analysis.

The $\mathcal{M}$ parameter measures the sphericity of the aggregate: if a cluster displays spherical symmetry, as in the case of a micelle, the value of the parameter $\mathcal{M}$ tends to 1.
With respect to $\mathcal{M}$, the $\mathcal{V}$ parameter is sensitive to the parallel or antiparallel alignment of the particles to the $\Vec{r}_{\mathrm{cm}} - \Vec{r}_i$ vector, which allows to distinguish vesicle-shaped clusters from micellar ones.
Finally, the $\mathcal{B}$ parameter quantifies the long range orientational parallel/antiparallel alignment within a cluster. In the case of a perfect bilayer, $\mathcal{B}$ tends to 1. 
The knowledge of $\mathcal{M}$, $\mathcal{V}$, $\mathcal{B}$ allows to uniquely classify the aggregates according to the thresholds reported in Table~\ref{tab:clust_ident}. Note that, as we focus on finite clusters, we label any aggregate with $N_c>N/2$ as ``extended cluster'' independently of the values of $\mathcal{M}$, $\mathcal{B}$ and $\mathcal{V}$.
As simulations are performed with $N = 2500$ particles, the threshold value to define an extended cluster is 1250.

\begin{table}[!h]
    \centering
    \begin{tabular}{|l|l|l|l|l|}
        \hline
        Cluster type (T) & $\mathcal{M}$ & $\mathcal{V}$ & $\mathcal{B}$ & $N_c$ \\
        \hline
        Micelle & $\geq 0.9$ & ... & ... & $< N/2$ \\
        Vesicle & $< 0.6$ & $> 0.5$ & ... & $< N/2$ \\
        Bilayer & $< 0.5$ &  ... & $\geq 0.4$ & $< N/2$ \\
        Elongated bilayer micelle & $\in [0.5, 0.9)$ & ... & $> 0.4$ & $< N/2$ \\
        Elongated micelle & $\in [0.5, 0.9)$ & ... & $< 0.4$ & $< N/2$ \\
        Extended cluster & ... & ... & ... & $> N/2$ \\
        \hline
    \end{tabular}
    \caption{Criteria for the cluster classification on the basis of $\mathcal{M}$, $\mathcal{V}$, $\mathcal{B}$ and $N_c$. } 
    \label{tab:clust_ident}
\end{table}

\begin{figure}[h]
    \centering
\includegraphics[width=1\columnwidth]{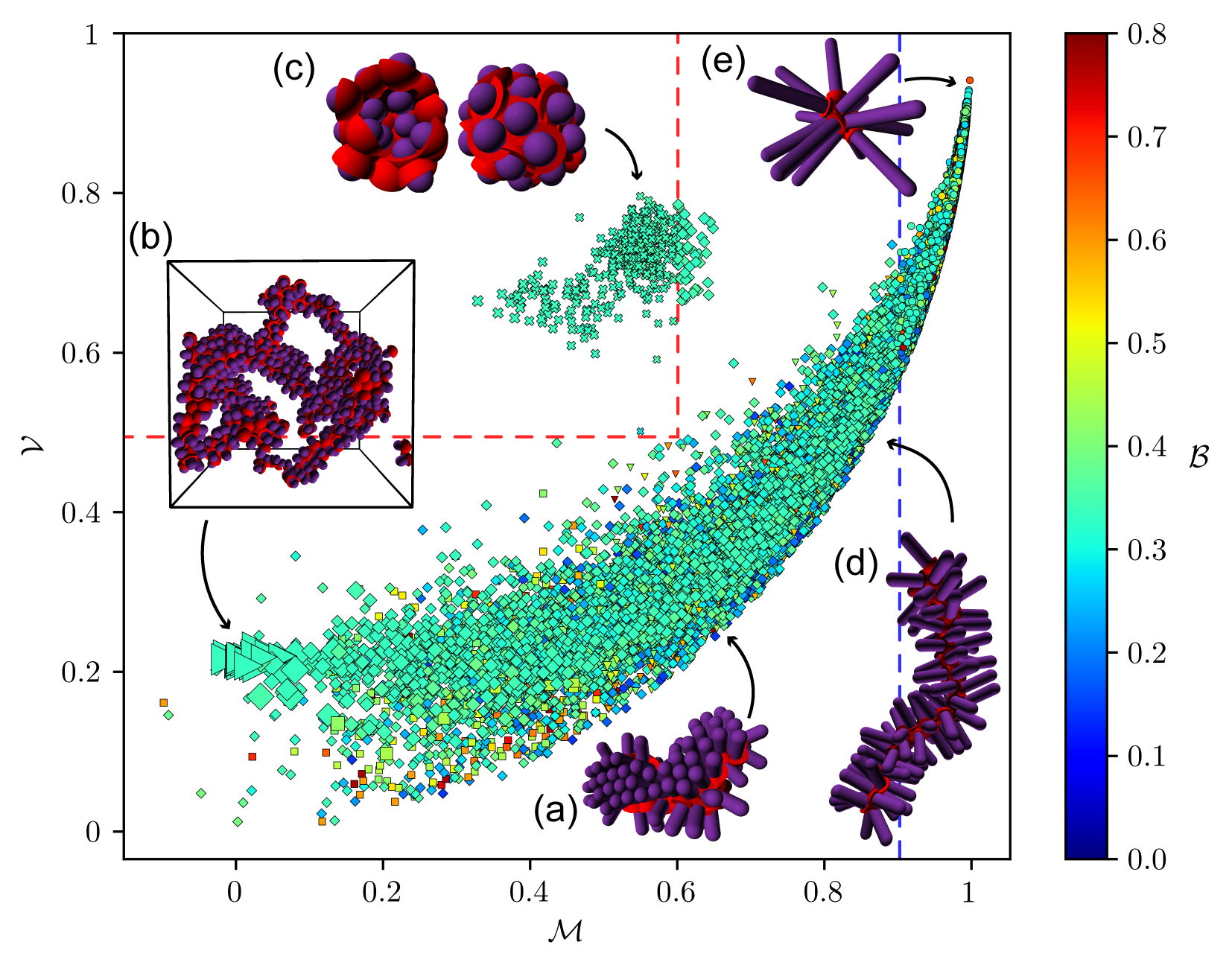}
\caption{Scatter plot in the $(\mathcal{M}, \mathcal{V})$ plane reporting the values computed for each cluster. A set of representative snapshots of clusters found within the simulations are reported: inset (a) elongated bilayer micelle, inset (b) extended structure, inset (c) vesicle (left view: vesicle cut by a plane to show its inner structure, right view: whole vesicle), inset (d) elongated micelle, inset (e) spherical micelle.  Different symbols identify the diverse clustered structures found within the simulations: right-pointing triangles identify the extended clusters, micelles are depicted as circles, crosses are representative of vesicles, elongated micelles are reported as diamonds, down-pointing triangles are the elongated bilayer micelles, and finally squares identify the bilayers. Each symbol representing a cluster is coloured accordingly to its $\mathcal{B}$ value, while the symbol size is a function of $N_c$. Dashed lines highlight the regions of the $(\mathcal{M}, \mathcal{V})$ plot where spherical micelles (blue line) and vesicles (red lines) can be found according to  Table~\ref{tab:clust_ident}.}
\label{fig:MVB}
\end{figure}

The graphical representation of the cluster classification is provided in Fig.~\ref{fig:MVB}, where every cluster identified within all simulations is represented according to the corresponding \{$\mathcal{M},\mathcal{V},\mathcal{B}$\} values.
In the $xy$-plane we report ($\mathcal{M}$, $\mathcal{V}$), while $\mathcal{B}$ is used to colour code the markers; the size of the markers is proportional to $N_c$. Following the classification reported in table \ref{tab:clust_ident}, the plot can be divided into different regions according to the values of the cluster order parameters. 
In particular, we highlight the region corresponding to spherical micelles ($\mathcal{M}>0.9$ and any value of $\mathcal{V}$ and $\mathcal{B}$) and report a snapshot of a typical aggregate (inset (e)).  
The plot clearly shows that, moving along a master curve on decreasing $\mathcal{M}$, the stable spherical micelles region is followed by an elongated micellar one, as exemplified by snapshots (d), corresponding to an elongated micelle, and (a), corresponding to an elongated small cluster. We also identify a very distinct region where vesicles prevail -- see snapshot (c) -- even though the distinction between vesicles and elongated micelles can become subtle at large values of $\mathcal{M}$. We finally observe that ``extended structures'' -- see snapshot (b) -- collect in the region where $\mathcal{M}\approx0$, $\mathcal{V}\approx0.2$, and $\mathcal{B}\approx 1/3$. As a matter of fact, irrespective of the actual shape of the extended cluster (being for instance a disordered fluid or an elongated and branched cluster) the latter values correspond to average over a set of random orientations. It can be noted that for the most of the formed clusters $\mathcal{B} \approx 1/3$, which implies that bilayers are extremely rare within our systems.

\begin{comment}{
Per confrontare con i dumbbell della Dijkstra, il caso sovrapponibile che non sia sfere è il caso  A=0.5 degli sferocilindri da paragonare con il suo l= 0.11/0.13 , considerando il range Delta=0.5 (ovvero il  delta/2= 0.25 di Carlo) e l'energia di bonding più alta (beta epsilon circa 4); in questo caso  otteniamo le stesse fasi, ovvero solo micelle elongate tra il caso sferocilindro e il caso dumbbell. Dijkstra ottiene le vescicole in quel caso solo andando ad aspect ratio leggermente più alti (attorno a l=0.16-0.2) dove però l'angolo del Kern Frenkel usando la loro formula risulta più elevato (nel range 93°-97° invece di 90° come per una Janus)}
\end{comment}

With the cluster analysis and related classification at hand, we define the probability that a system aggregates into a cluster of kind K as:
\begin{equation}
    P_\mathrm{K} = \sum_i^{n_\mathrm{K}} \frac{N_c^i}{N_c^{tot}}
\end{equation}
where $N_c^{tot}$ is the total number of clustered particles in the system, $n_\mathrm{K}$ is the total number of clusters of kind K at a given state point, and the sum runs over the $n_\mathrm{K}$ clusters of kind K, each of size $N_c^i$. On this basis, we assign to each system a dominant microphase K according to the greatest $P_\mathrm{K}$ value. Note that, if the fraction of clustered particles in the system ($N_c^{tot} / N$) is less than $0.5$, then the system is considered to be below the critical micelle concentration (cmc). 

\begin{figure*}[!ht]
    \centering
\includegraphics[width=1\textwidth]{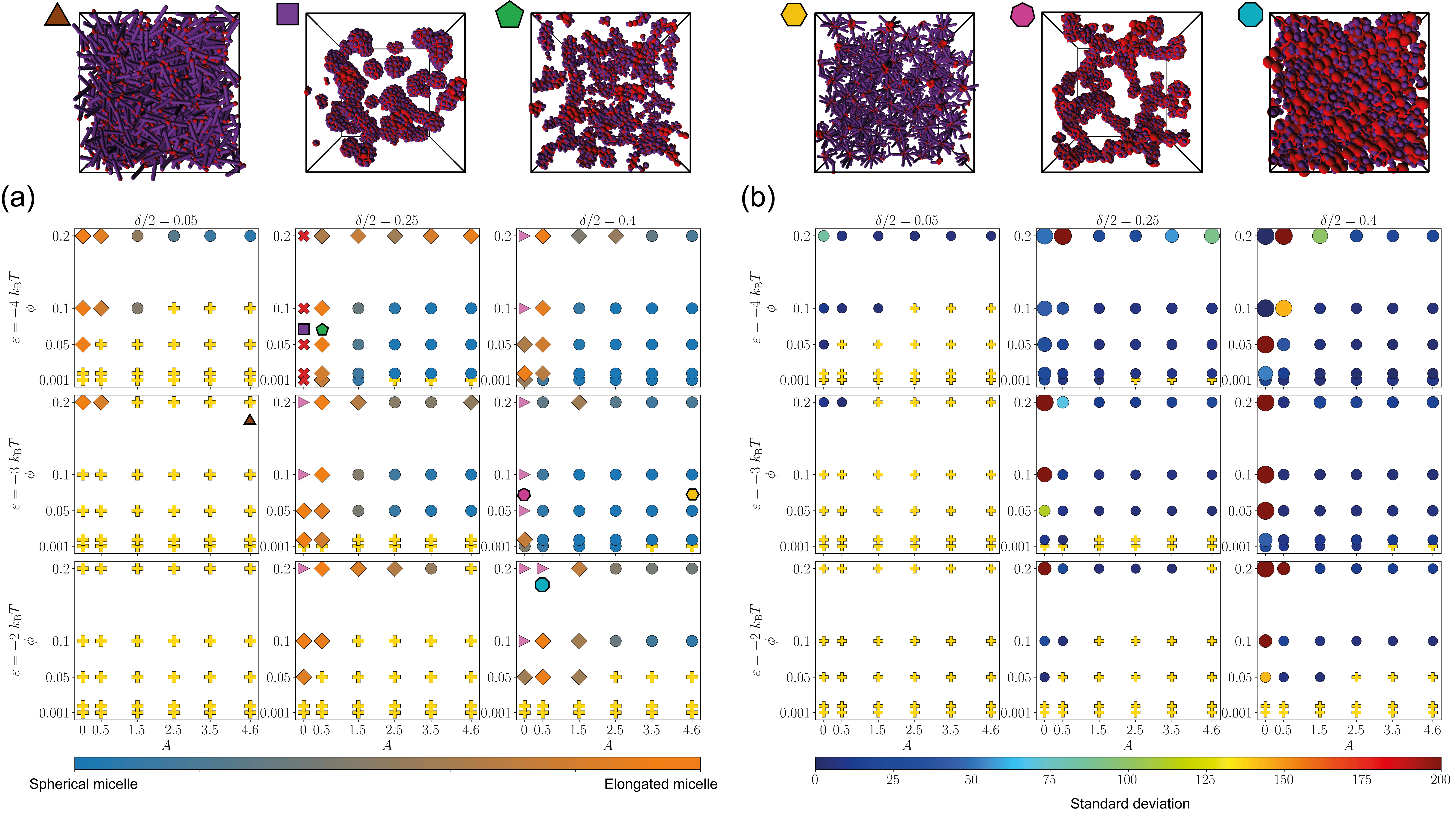}
\caption{Panel (a): phase diagram as a function of $A, \delta, \epsilon$ and $\phi$ reporting the most probable cluster phase for every state point. The marker type represents the dominant cluster phase: systems below cmc (yellow plus symbols), vesicles (red crosses), extended clusters (pink triangles), spherical micelles (circles), and elongated micelles (diamonds). About micelles, to highlight the coexistence of spherical and elongated structures, the corresponding marker colour reflects the cluster type probability, spanning from blue for predominantly spherical micelles to orange for predominating elongated ones. Panel (b): mean (marker size) and standard deviation (marker colour) of the cluster size distribution for each system.  For sake of clarity, we associate  illustrative snapshots of typical configurations to the symbols used in the phase diagram; brown triangles represent system of rods under the cmc; purple square: vesicles; green pentagons: elongated micelles; yellow hexagons: spherical micelles; pink heptagons: extended cluster formed by a bent tube; teal octagons: liquid-like extended cluster.}
\label{fig:Phase_diagram}
\end{figure*}

The complete equilibrium phase diagram, made by the most probable microphase for each state point, is reported in panel (a) of Fig.~\ref{fig:Phase_diagram} as a function of $A$, $\delta$, $\varepsilon$, and $\phi$. For the same state points, panel (b) shows the average cluster size and the standard deviation, irrespective of the cluster type. 

The emergence of microscopic structures is strongly influenced by the interaction strength, its range, and the packing fraction of the system.  As a matter of fact, at the lowest values of $(\delta/2, \varepsilon)$ (see bottom-left of panel (a) in Fig. \ref{fig:Phase_diagram}) all the systems are below the cmc for all elongations and densities. For the smallest interaction range ($\delta / 2 = 0.05$) the cmc is reached almost exclusively when the interaction strength is the highest, i.e. at $\varepsilon = -4 k_\mathrm{B} T$ (top-left of panel (a)). At this value of $\varepsilon$, the cmc shifts towards higher $\phi$ values on increasing $A$, since systems composed by more elongated particles have a lower effective patch density for a given packing fraction (brown-triangle snapshot in Fig.~\ref{fig:Phase_diagram}). The whole phase diagram of the systems as a function of  $\rho = N/V$ is reported in Section 2 of the SI. 
The cmc shows a non-trivial dependence on the $(A,\delta,\varepsilon)$ combination. In fact, for $\delta / 2 = 0.25$ and $0.4$ (central and right rows of panel (a) in Fig.~\ref{fig:Phase_diagram}) the cmc is reached already at intermediate/low interaction strengths, namely at $\varepsilon = -3 k_\mathrm{B} T$ and $-2 k_\mathrm{B} T$, respectively. Nonetheless, also for $\delta / 2 = 0.25$ and $0.4$, the $\phi$ values at which the cmc appears  increase on increasing $A$ as a consequence of the reduced effective density of the patches at larger particle elongations.

By reading the phase diagram as a function of the particle anisotropy $A$, it is possible to analyse separately the effects on the aggregates due to the entropic and the enthalpic contributions. 

The $A = 0$ case represents Janus spheres, which have been widely analysed in the literature \cite{Sciortino2009, Sciortino2010, Sciortino2013, Sciortino2014, Vissers2013, Sciortino2016, Sato2020, Li2012, Zou2016, Diaz2019} and present a very rich phase diagram. For the $(\phi, \delta, \varepsilon)$ combinations analysed in this study, systems are keen to cluster in elongated micelles or extended aggregates, while spherical micelles are unlikely to form and vesicles appear only for $(\delta/2,\varepsilon)=(0.25,-4  k_\mathrm{B} T)$ (top-centre of panel (a), purple-square snapshot), in  line with what observed in Ref.~\cite{Sciortino2009, Sciortino2010} (see Section 2 of SI). In general, the higher the packing fraction, the higher the probability of forming extended clusters. As reported in panel (b) of Fig. \ref{fig:Phase_diagram}, the average cluster size -- represented by the size of the symbols -- increases with $\phi$ at any combination ($\delta/2,\varepsilon$) above the cmc. These extended structures may be due  either to the merging of multiple elongated micelles/bilayers (as in the case of $\delta/2 = 0.4$ and $\varepsilon = (-3, -4) k_\mathrm{B} T$, pink-heptagon snapshot) or to the formation of a disordered phase where the particles inside the cluster do not show a preferential bonding orientational order (as in the cases of $\delta/2 = 0.25$ and $\varepsilon = (-2, -3)k_\mathrm{B} T$ or $\delta/2 = 0.4$ and $\varepsilon = -2 k_\mathrm{B} T$), see  Fig. 3 in Section 4 of SI for a  comparison between the diverse intra-cluster arrangements.
Most of the extended clusters coexist with smaller structures, resulting in a high standard deviation, as shown by the colors of the symbols in panel (b) of Fig. \ref{fig:Phase_diagram}.

The effect of a relatively small anisotropy ($A=0.5$) on the described cluster phases varies with the combination of ($\delta/2,\varepsilon$) values. Most ($\delta/2,\varepsilon$) combinations still lead to extended clusters or micelles, whereas the most significant morphological changes are observed for $(\delta/2,\varepsilon)=(0.25,-4 k_\mathrm{B} T)$ and $(\delta/2,\varepsilon)=(0.4,-3 k_\mathrm{B} T)$. In the first case, 
the increase of steric hindrance  inhibits the formation of stable vesicles. 
This behaviour is in line with what observed by Avvisati \textit{et al.} in \cite{Dijkstra2015}, where the authors studied the phase diagram of systems of patchy dumbbells as a function of their size ratio $q$ and the distance $l$ between the spheres. These systems can be compared to our tip-functionalised rods when the parameters of the potential are similar to each other (meaning $\delta/2 = 0.25$ and $-4 k_\mathrm{B} T < \varepsilon < -3 k_\mathrm{B} T$) and the Janus-like opening angle of the patches is guaranteed (meaning $\theta = \pi/2$). Among the cases studied by the authors, the case of $q = 1.035$ and $l \approx 0.13$ corresponds to a Janus-like rod with anisotropy $0 < A < 0.5$. Analogously to our results for the same $\varepsilon$ and $\delta$, the authors observed that the introduction of even a small anisotropy suppresses the vesicular phase while stabilising elongated micelles. It is interesting to notice that according to Ref.~\cite{Dijkstra2015} an opening angle greater than $\pi/2$ of the K-F potential is required to recover a vesicular phase for these systems. 
It is worth mentioning that the emergence of elongated micelles (green-pentagon snapshot of Fig.~\ref{fig:Phase_diagram}) observed for $(\delta/2,\varepsilon)=(0.25,-4 k_\mathrm{B} T)$ also occurs at lower $\varepsilon$ for the same $\delta$. At this value of $\delta$, the elongated structures become characterised by an increasing number of bonds on increasing $\varepsilon$ 
(Fig. 4 in Section 4 of SI): a high enthalpic contribution in fact allows to balance the concurrent decrease of entropy of the system which is more relevant as the aggregates become bigger and/or more compact.
Also for the highest interaction range ($\delta/2 = 0.4$), it is the interaction strength $\varepsilon$ that controls the equilibrium between the enthalpic and the entropic contributions as it renders the formation of a bond a major driving force irrespective of the entropy loss. In fact, on increasing $\varepsilon$, clusters with a higher number of bonds per particle appear  (Fig. 5 in Section 4 of SI). 
The other combination of parameters showing a striking morphological difference with respect to the spherical case is $(\delta/2,\varepsilon)=(0.4,-3 k_\mathrm{B} T)$. 
For this combination of energy strength and interaction range, spherical micelles emerge at $A=0.5$, which were completely absent in the phase diagram of the Janus spheres (center-right of panel (a) of Fig. \ref{fig:Phase_diagram}). 

When $A>0.5$, spherical micelles become the most probable aggregates, for a wide range of packing fractions at intermediate and large $\varepsilon$ values. 
As a matter of fact, the introduction of a shape anisotropy beyond $A=0.5$ strengthens the repulsive effect due to the excluded volume. Particles tend to maximise the average distance between the hard cores while maximising the contacts between the attractive tips, thus assembling into a spherical micelle. For low to intermediate $\phi$ values, till micelles start to interact with one another, elongated cluster structures are thus disfavoured  in favour of the formation of a gas of spherical micelles, that allows both for the minimisation of the interaction between the hard part of the rods, as well as for the maximisation of the configurational entropy.
Upon increasing $\phi$ (see in particular $\delta/2 = 0.25$ and $\varepsilon = -4 k_\mathrm{B} T$) the average distance between spherical clusters decreases. This forces particles to interact, leading to the formation of elongated clusters with aligned particles, arising from the coalescence of spherical micelles.
As a consequence, there are regions in the phase diagram where elongated and spherical micelles coexist. When this happens, the predominant phase is assigned according to the largest value of the corresponding probability $P_\mathrm{T}$. 
This behavior let us envisage that for higher elongations, the phase transition between spherical and elongated micelles shifts towards lower packing fractions, following the trend of the I-N transition for unfunctionalised hard spherocylinders \cite{Pal2022,CADF}

As for the average size of the cluster,  an increase in  $\delta$, $\varepsilon$ or $\phi$ (individually  or in combination) lead to the formation of bigger clusters  as shown in  Fig. \ref{fig:Phase_diagram} panel (b). 
On the contrary, an increase in $A$, enhances the  steric hindrance between the particles, thus disfavouring the formation of large clusters.
Moreover, all spherical micelles appear to be quite monodisperse, and the degree of polydispersity decreases upon increasing $A$.

Finally, it is interesting to observe that although some spare bilayer clusters are seen in our simulations (see Fig. \ref{fig:MVB}), stable bilayers or elongated micelle bilayers are never predominant.

\section{Conclusions}

In this study we considered Janus-like spherocylinders and performed Monte Carlo simulations with Aggregation-Volume-Bias in the NVT ensemble to extensively investigate the aggregation behaviour of this type of particles on systematically varying strength and interaction range of the attractive patch. We focused in particular on the low density regime, where finite clusters are expected, to separately assess the role played by particle shape anisotropy and tip functionalisation on the cluster morphology. 

Our analysis highlights a strong dependence of the average cluster shape on the particle shape anisotropy. We show that spherical Janus particles never stabilise spherical micelles and rather form elongated micelles, vesicles and extended clusters. As soon as anisotropy is introduced in the colloidal constituents, the sphericity of the clusters is recovered: spherical micelles are thus stabilised for a wide range of parameters in the phase space, even though it is worth noting that for the highest packing fractions analysed, spherical micelles merge into elongated micelles. Additionally, vesicles -- that for spherical Janus colloids are stable for a narrow region of strength and interaction range -- disappear as soon as particles become elongated. We finally observed that increasing the aspect ratio of the particles inhibits the formation of extended clusters.

A striking feature of the emerging micellar phases is the enhanced monodispersity of these spherical clusters for any combinations of shape anisotropy, interaction range and interaction strength at which they emerge. The ensemble of the collected results thus suggests that shape anisotropy can be exploited to stabilize spherical micelles of Janus colloids with respect to other cluster types, thus guaranteeing a better control on those features of the aggergates that are relevant for application purposes.

\section{Suuporting Information} 
\subsection{Aggregation-Volume-Bias}\label{SI:AVB}
The Aggregation Volume Bias Monte Carlo (AVB-MC) algorithm is designed to improve the phase space sampling  of cluster forming systems. AVB-MC enhances the acceptance rate of trial moves that result in molecular clusters breaking up, and simultaneously increases the transition probability of moves that lead to the formation of bonded structures \cite{doi:10.1021/jp001952u, Chen_AVB, Chen_AVB_improved}.
In this work  the AVB-MC algorithm is designed so that the system has a 90\% probability of making either a translation or a rotation move, where both moves are chosen to have equal probability. To improve the phase space sampling, we introduce a biased move with a probability of 10\%, which can be either a bonding (AVB-B) or an unbonding (AVB-U) trial move with equal probability. The AVB-B move is designed to form a bond between two unbounded particles, while the AVB-U move is designed to break an existing bond. The standard AVB-MC algorithm, makes use of the  bonding volume  $V_b$ -- defined  as the portion of the phase-space in which two particles are bonded -- to weight both AVB-B and AVB-U moves within the metropolis algorithm.
In the case of two particles of diameter $D$ decorated with a Kernel-Frenkel patch of $\delta/2$ range and of $\theta$ half opening, the bonding volume can be computed analytically as:

\begin{equation}\label{eq:Vbond}
    V_b = \frac{4 \pi^2}{3} (1 - \cos \theta)^2 \left ( \left (D + \delta \right )^3 - D^3 \right )
\end{equation}

Let us introduce the concept of  outer volume $V_o$, defined  as the number of non-bonding configurations in the phase space:
    
\begin{equation}\label{eq:Vout}
    V_o = 4 \pi V - V_b
\end{equation}

where $V$ is the volume of the simulation box and $4 \pi$ comes from the orientational degrees of freedom.
Given a system of $N_p$ patches, where $N_i$ is the number of patches bound to the $p_i$ patch of the $i$-th particle, then within the AVB-MC scheme:

\begin{enumerate}
    \item in the case of the AVB-B move, the trial is accepted with probability
    \begin{equation}\label{eq:AVB_B_acc}
        acc(o \to n) = min \left\{1, \frac{(N_p - N_i - 1) V_b}{(N_i + 1) V_o} e^{- \beta \Delta \mathcal{U}} \right\}
    \end{equation}
    where $\Delta \mathcal{U}$ is the energy difference between the two configurations

\item 
while in the case of AVB-U move, the acceptance probability is

\begin{equation}\label{eq:AVB_U_acc}
    acc(o \to n) = min \left\{1, \frac{N_i V_o}{(N_p - N_i) V_b} e^{- \beta \Delta \mathcal{U}} \right\}
\end{equation}
\end{enumerate}

AVB-MC algorithm has been used so far in literature to enhance the sampling in the case of cluster forming spherical particles. In this work, the algorithm is generalised to the case of elongated rods; this requires that, once the patch is moved  into a bonded/unbonded configuration, the body of the elongated particle has to be rototranslated to rebuild the original symmetry of the particle.

\subsection{Phase diagram as a function of $\rho$}

\begin{figure}[h]
    \centering
\includegraphics[width=1\columnwidth]{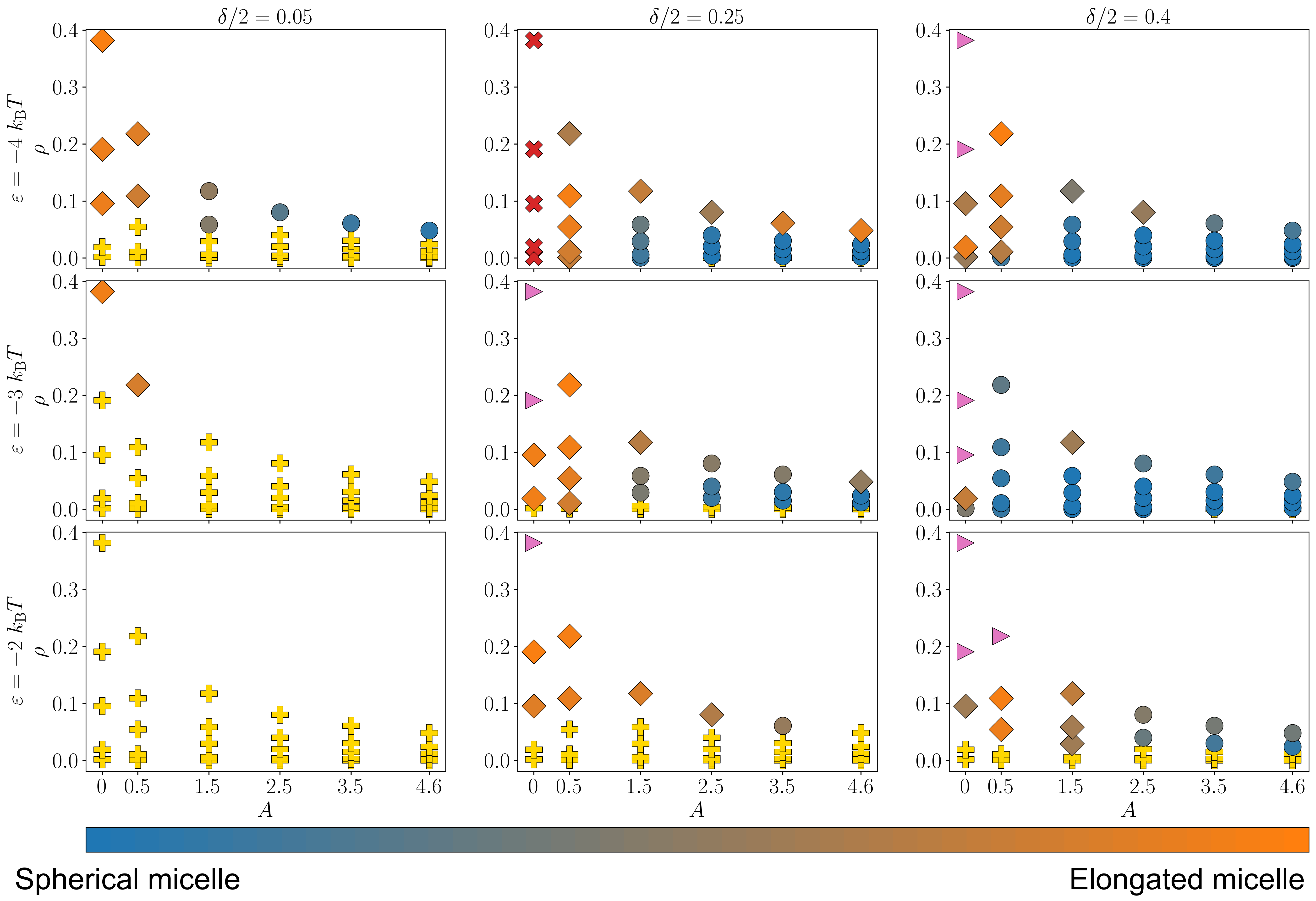}
    \caption{Phase diagram as a function of $A, \delta, \varepsilon$ and $\rho = N/V$ reporting the most probable cluster phase for every state point. The marker type represents the dominant cluster phase: systems below cmc (yellow plus symbols), vesicles (red crosses), extended clusters (pink triangles), spherical micelles (circles), and elongated micelles (diamonds). About micelles, to highlight the coexistence of spherical and elongated structures, the corresponding marker colour reflects the cluster type probability, spanning from blue for predominantly spherical micelles to orange for predominating elongated ones.}
    \label{SI:fig:phase_diagram_rho}
\end{figure}

In Figure \ref{SI:fig:phase_diagram_rho} we report the phase diagram of the system as a function of the density $\rho = N/V$. It is possible to appreciate that the cmc for fixed ($\delta$, $\varepsilon$) is reached almost at the same density for all the elongations $A$. This is due to the fact that the emergence of clustering only depends on the tip-tip interactions, that is the effective density of patches.

\subsection{Literature comparison}

\begin{figure}[h]
    \centering
\includegraphics[width=0.8\columnwidth]{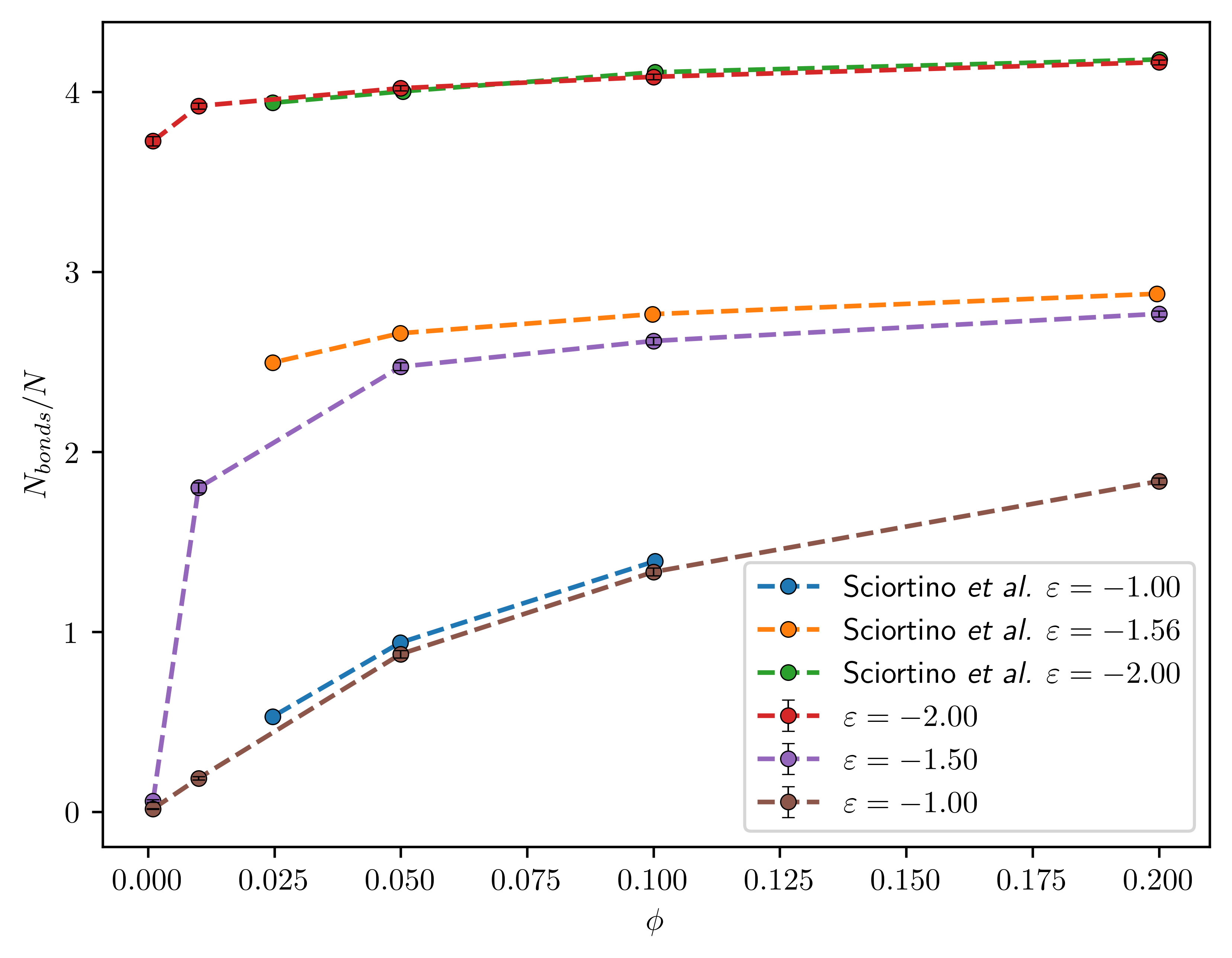}
    \caption{Average number of bonds per particle for Janus spheres ($A = 0$) as a function of $\phi$ and $\varepsilon$ for systems with $\delta/2 = 0.25$. Data from Sciortino \textit{et al.} \cite{Sciortino2009, Sciortino2010}  (blue, orange and green symbols) are reported along with our data (red, purple and brown symbols) for comparisons.}
    \label{SI:fig:Sciortino}
\end{figure}

We compared the average number of bonds per particles as a function of $\phi$ and $\varepsilon$ that we obtained for Janus spheres ($A = 0$) with interaction range $\delta/2 = 0.25$ with available data from literature (Sciortino \textit{et al.} \textcolor{red}{\cite{Sciortino2009, Sciortino2010}}). The comparison is reported in Fig. \ref{SI:fig:Sciortino}, which shows the very good agreement between the two data sets.

\subsection{Analysis of intra-cluster arrangement}

The three order parameters used in this work ($\mathcal{M}$, $\mathcal{V}$, and $\mathcal{B}$) describe the overall structure of the cluster.
To have a more in-depth understanding of what differentiates the different cluster structures at fixed state points in the phase diagram, we set our attention on the relative orientations between all the bonded couples of particles. To this aim, we  computed the distribution of scalar products $P(\hat{n}_i \cdot \hat{n}_j)$, where $\hat{n}_i$ and $\hat{n}_j$ are the unit vectors describing the orientation of particles $i$ and $j$. Moreover, we obtained the average number of bonds per particle.

\begin{figure}[h]
    \centering
\includegraphics[width=1\columnwidth]{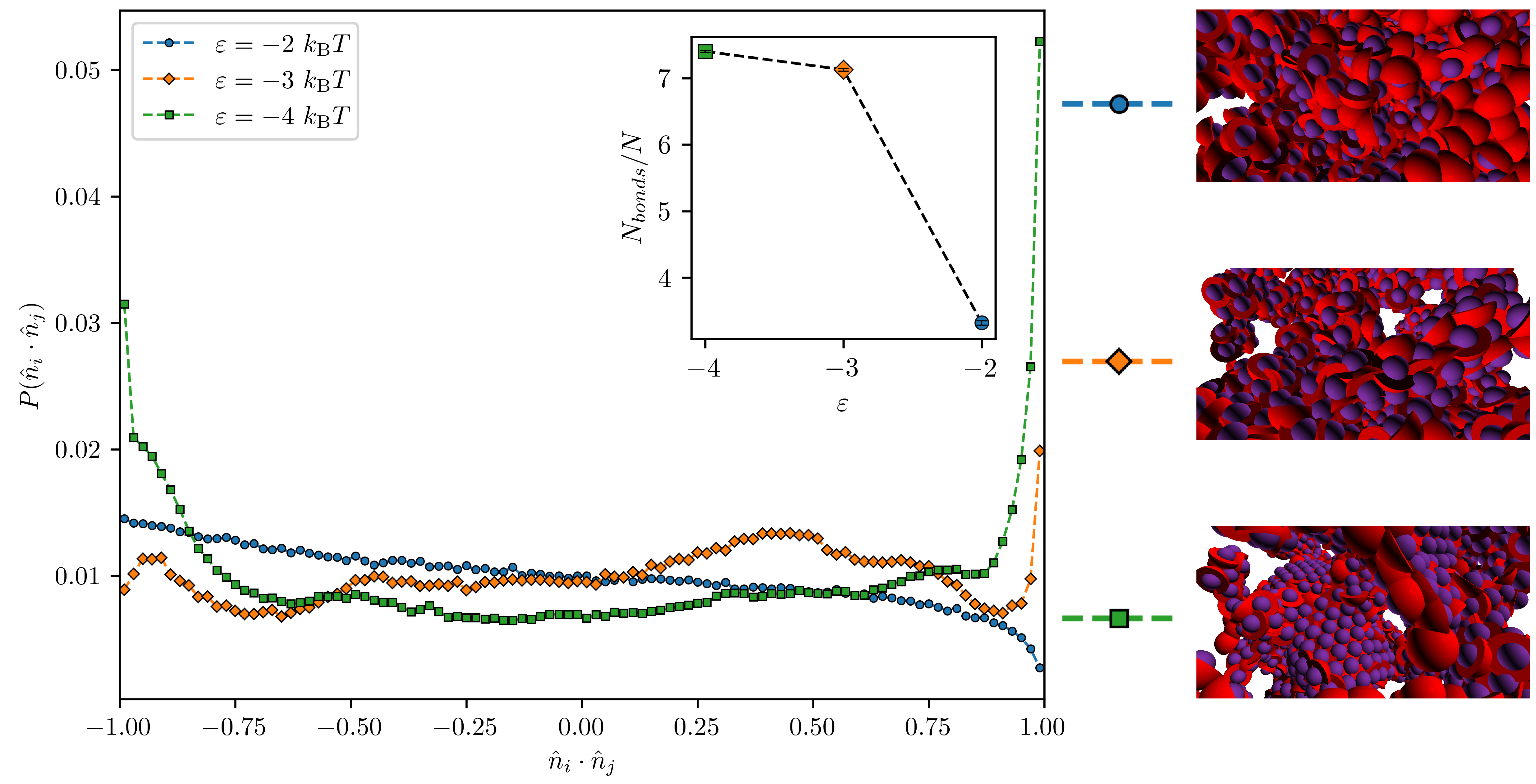}
    \caption{Distribution of the scalar product $\hat{n}_i \cdot \hat{n}_j$ evaluated over all pairs of bonded particles for the system with $\delta / 2  = 0.4$, $A = 0$, and $\phi = 0.2$ as a function of $\varepsilon$. In the inset the average number of bonds per particle is reported. Representative snapshots of the three types of extended clusters analysed are also included.}
    \label{SI:fig:Extended_cluster_bondsprod}
\end{figure}

In Fig. \ref{SI:fig:Extended_cluster_bondsprod} we analysed the extended clusters emerging from systems with $\delta / 2  = 0.4$, $A = 0$, and $\phi = 0.2$ as a function of $\varepsilon$. It is clear that for $\varepsilon = -2 k_\mathrm{B}T$ the particles arrange into a disordered liquid-like structure. The ordering increases with $\varepsilon$, leading to a concurrent increase in the number of bonds.

\begin{figure}[h]
    \centering
\includegraphics[width=1\columnwidth]{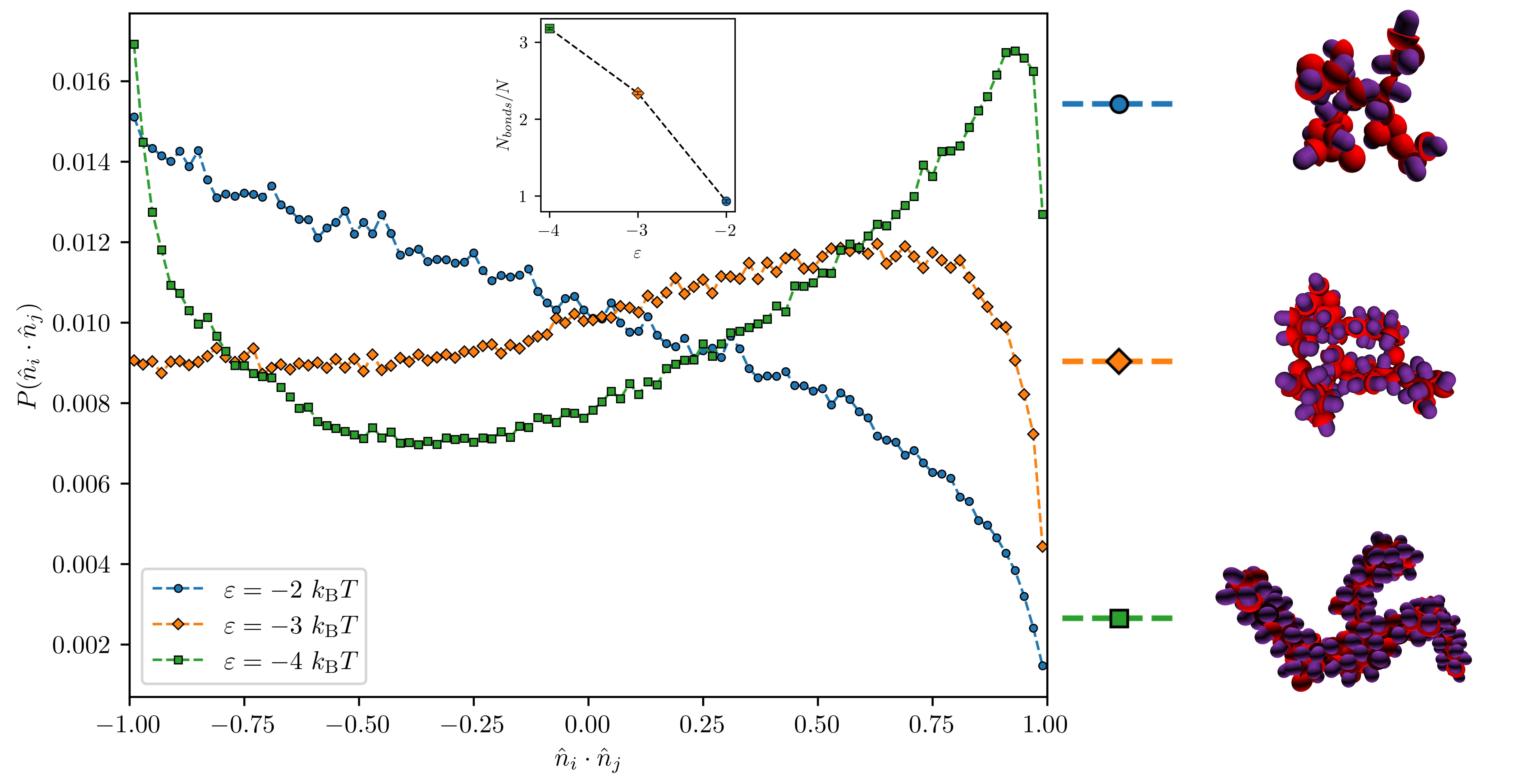}
    \caption{Distribution of the scalar product $\hat{n}_i \cdot \hat{n}_j$ evaluated over all pairs of bonded particles for the system with $\delta / 2  = 0.25$, $A = 0.5$, and $\phi = 0.1$ as a function of $\varepsilon$. In the inset the average number of bonds per particle is reported. Representative snapshots of the cluster with the maximum size observed in the systems analysed are also included.}
    \label{SI:fig:elongated_bondsprod}
\end{figure}

The nature of the elongated micelles emerging in the systems with $\delta / 2  = 0.25$, $A = 0.5$, and $\phi = 0.1$ as a function of $\varepsilon$ is analysed in Fig. \ref{SI:fig:elongated_bondsprod}. With $\varepsilon$ the internal order of the structures increases, which is mirrored both by an increase in the average number of bonds and by a transition from a disordered structure ($\varepsilon = - 2 k_\mathrm{B}T$) to a more ordered one ($\varepsilon = - 4 k_\mathrm{B}T$).

Finally, we analysed the apparent reentrant behaviour as a function of $\varepsilon$ that is observed for systems with $\delta / 2  = 0.4$, $A = 0.5$, and $\phi = 0.1$. 
In fact, for this system,  upon augmenting the interaction strength $\varepsilon$, the most stable phases are elongated micelles for $\varepsilon = - 2 k_\mathrm{B}T$, spherical micelles for $\varepsilon = - 3 k_\mathrm{B}T$, and again elongated micelles for $\varepsilon = - 4 k_\mathrm{B}T$. To understand such a feature, we can focus on the local arrangement of particles in the different clusters, in terms of the relative orientation of particles. Through the combination of the distributions $P(\hat{n}_i \cdot \hat{n}_j)$ and the average number of bonds per particle, it emerges that:
\begin{itemize}
\item in the case of $\varepsilon = -2 k_\mathrm{B} T$ the particles arrange into a liquid-like structure;
\item  for $\varepsilon = -3  k_\mathrm{B} T$ the spherical nature of the clusters emerges;
\item for $\varepsilon = -4  k_\mathrm{B} T$ the particles form elongated clusters, locally resembling bilayers and/or long tubes.
\end{itemize}
This indicates that, even if the system apparently stabilises twice elongated clusters, the intrinsic nature of the aggregate is different. For the lowest value of $\varepsilon$ elongated clusters are disordered, differently than the ones assembled at the highest value of $\varepsilon$ that, arising form the merging of spherical micelles, present a local order between the colloidal particles. 

\begin{figure}[h]
    \centering
\includegraphics[width=1\columnwidth]{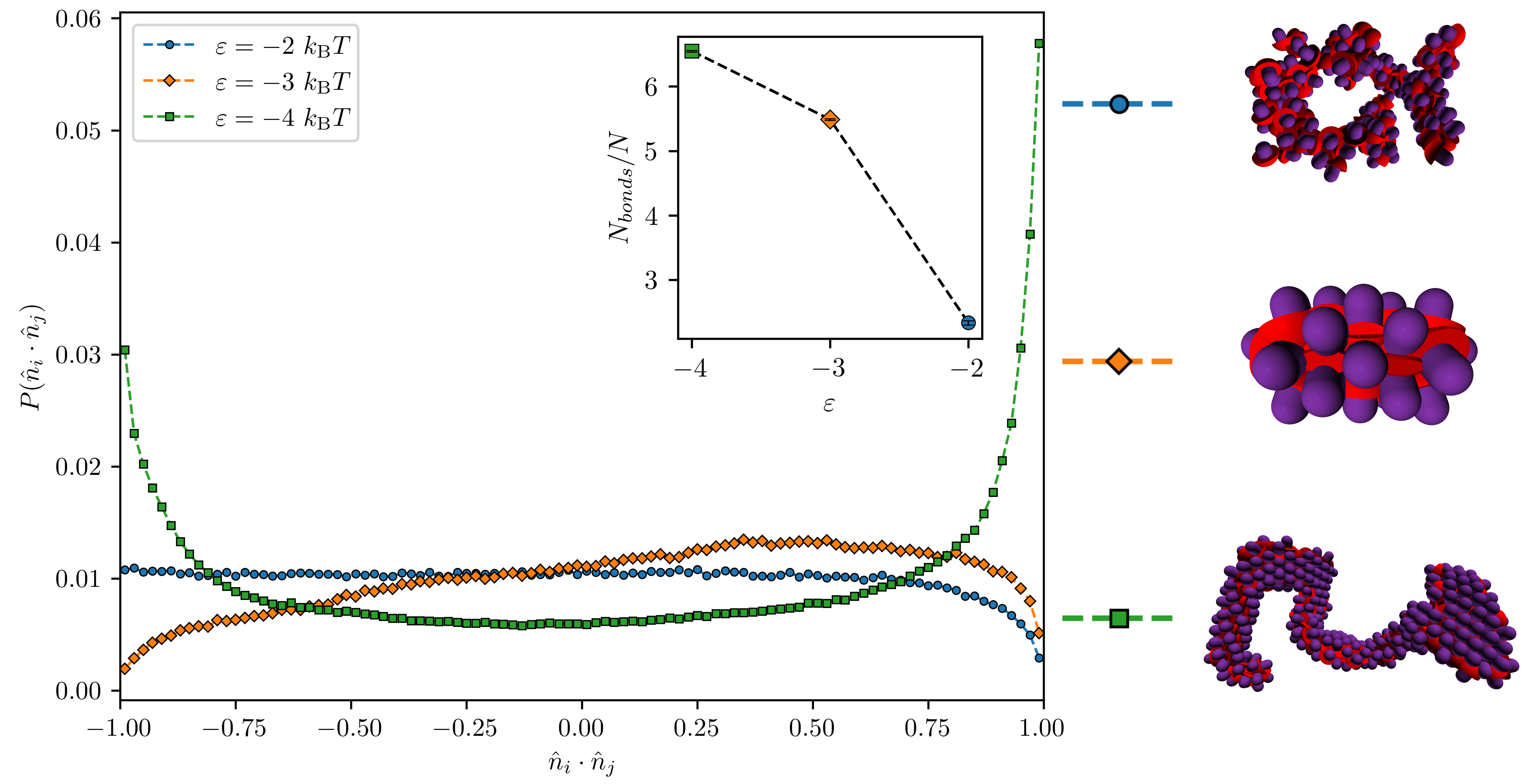}
    \caption{Distribution of the scalar product $\hat{n}_i \cdot \hat{n}_j$ evaluated over all pairs of bonded particles for the system with $\delta / 2  = 0.4$, $A = 0.5$, and $\phi = 0.1$ as a function of $\varepsilon$. In the inset the average number of bonds per particle is reported. Representative snapshots of the cluster with the maximum size observed in the systems analysed are also included.}
    \label{SI:fig:elongated_to_spherical_bondsprod}
\end{figure}

\section*{Acknowledgments} 
CADF, SDG and BC would like to acknowledge the Grant of Excellence Departments, MIUR-Italy (ARTICOLO 1, COMMI 314 - 337 LEGGE 232/2016). and the Rome Technopole Project (CUP:F83B22000040006). SDG and BC acknowledge the funding  PON ``Ricerca e Innovazione'' 2014-2020 D.M. 1423. 16-09-2022. CADF and BC acknowledges financial support from European Union - Next Generation EU (MUR-PRIN2022 PRIN 2022RYP9YT SCOPE  CUP:F53D23001130006).  CDM acknowledges financial support from European Union - Next Generation EU (MUR-PRIN2022 TAMeQUAD CUP:B53D23004500006).
EB acknowledges support from the Austrian Science Fund (FWF) under Proj. No. Y-1163-N27.

%\bibliography{bibliography.bib}
%\bibliographystyle{rsc}

\providecommand*{\mcitethebibliography}{\thebibliography}
\csname @ifundefined\endcsname{endmcitethebibliography}
{\let\endmcitethebibliography\endthebibliography}{}

\end{document}